\documentclass[preprint,12pt,3p]{elsarticle}

\usepackage{amssymb}

\begin{document}

\begin{frontmatter}

\title{High energy density spots and production of kink-antikink pairs in particle collisions}

\author[label1]{Alidad Askari}
\address[label1]{Department of Physics, Ferdowsi University of Mashhad,
91775-1436 Mashhad, Iran}
\ead{al.askari@stu.um.ac.ir}
\author[label2]{Danial Saadatmand}
\address[label2]{Department of Physics, University of Sistan and Baluchestan, Zahedan, Iran}
\ead{saadatmand.d@gmail.com}
\author[label3]{Sergey V. Dmitriev\fnref{label4}}
\address[label3]{Institute for Metals Superplasticity Problems RAS, Khalturin Street 39, 450001 Ufa , Russia}
\address[label4]{Research Laboratory for Mechanics of New Nanomaterials, Peter the Great St. Petersburg Polytechnical University, St. Petersburg 195251,
Russia}
\ead{dmitriev.sergey.v@gmail.com}
\author[label1]{Kurosh Javidan\corref{cor1}}
\ead{javidan.um.ac.ir}
\cortext[cor1]{I am corresponding author}

%
%
%
%
%
%

\begin{abstract}
The maximal energy density that can be achieved in the collisions of the particle-like wave trains in the $\phi^4$ model has been investigated numerically for different wave train parameters. From these results the prediction is made on how many kink-antikink pairs can be produced in wave train collisions taking into account that in collision of $N$ kinks maximal energy density is equal to $N^2/2$ for even $N$. Our numerical results are in a reasonably good agreement with the predicted maximal number of kink-antikink pairs formed in wave train collisions.
\end{abstract}


\end{frontmatter}

\section{Introduction}

Solitons or solitary waves are ubiquitous in many branches of physics where nonlinearity plays an important role \cite{Book1,Book2,BraunKivshar,Book3}. Solitons are robust with respect to small perturbations, they emerge unchanged from collisions with each other, and they can travel long distances transporting energy in a concentrated form. In some cases, for instance, in the nonlinear Schr\"odinger equation, solitons can have any energy from zero to a large value \cite{Book2}. On the other hand, creation of topological solitons, e.g., in Klein-Gordon fields, requires relatively large amount of energy greater than a threshold value \cite{BraunKivshar,Book3}. In any case the mechanisms of soliton excitation or energy pumping are very important and they have been discussed in a number of works. The probability of the topological soliton-antisoliton pair production in particle collisions has been calculated by Demidov and Levkov \cite{Demidov1,Demidov2}. Production of solitons from the scattering of small breathers has been analyzed for a wide range of initial conditions by Lamm and Vachaspati \cite{Lamm}. Kink-antikink pairs can be created in near-separatrix collisions of low-frequency (i.e. high-energy) breathers in weakly perturbed sine-Gordon field \cite{DKS2001,DKK2008}. A breather or an oscillon can split into a kink-antikink pair by a sudden distortion of the sine-Gordon potential into a $\phi^4$ potential \cite{CP2009}. The interaction between kink and radiation in $\phi^4$ model has been investigated and the role of these radiation in the creation of kink-antikink pairs has been discussed \cite{Romanczukiewicz}. Soliton creation induced by collisions of a few highly energetic particles has been analyzed \cite{Levkov}. It has been shown that production of the kink-antikink pairs from the collision of small-amplitude breather trains preferably occurs for lower incoming velocity \cite{Dutta}. In a later similar work the three steps in the process of production of kink-antikink pairs in the collision of particlelike states in the one-dimensional $\phi^4$ model have been revealed \cite{RomanczukiewiczPRL}.

A necessary condition for creation of a kink-antikink pair in a Klein-Gordon field is that a sufficiently high energy density spot should be achieved by any means. In our recent works we have analysed how large can be energy density in multi-soliton collisions in the sine-Gordon and $\phi^4$ models \cite{DanialPRD,Aliakbar}. In relation to the problem of kink-antikink pair formation it is tempting to analyze if there is a correlation between the value of the energy density produced by the colliding particles considered in \cite{RomanczukiewiczPRL} and the number of emerging kink-antikink pairs. In this work we address this issue for the $\phi^4$ model.

In Sec.~\ref{Sec:II} the problem to solve and the numerical scheme used in the simulations are described. In Sec.~\ref{Sec:III}, by integrating numerically the $\phi^4$ equation of motion, we estimate the maximal energy density observed in the collision of particles and demonstrate that from this result the number of the emerging kink-antikink pairs can be predicted. Our conclusions are given in Sec.~\ref{Sec:IV}.

\section {The model} \label{Sec:II}

The equation of motion of the $\phi^4$ model in (1+1) dimensions is
\begin{equation}\label{SG}
\phi_{tt} - \phi_{xx}-2\phi (1-\phi^2)= 0,
\end{equation}
where $\phi_{tt}$  and $\phi_{xx}$ are the second order derivatives of $\phi(x,t)$ scalar field with respect to the time and space, respectively. This equation has an exact solution which is known as moving kink/antikink
\begin{equation}\label{Kink}
\phi_{k\bar k}(x,t) = \pm\tanh{[\delta(x-Vt)]},
\end{equation}
where $V$ is the kink velocity and $\delta =1/\sqrt{1 -V^2}$. Positive (negative) sign in Eq.~(\ref{Kink}) is related to the kink (antikink).

The total energy of the scalar field is defined as
\begin{equation}\label{Energy}
U=\frac{1}{2}\int\limits_{-\infty}^{\infty} \Big[ \phi_t^2 +\phi_x^2 +(1-\phi^2)^2 \Big]dx\,.
\end{equation}
The corresponding integrand describes the total energy density of the $\phi^4$ field,
\begin{eqnarray}\label{Edensity}
u(x,t)=\frac{1}{2}\phi_t^2+ \frac{1}{2}\phi_x^2+\frac{1}{2}(1-\phi^2)^2.
\end{eqnarray}

For the $\phi^4$ model it has been shown that the maximal energy density that can be achieved in collision of $N$ slow kinks/antikinks  is equal to (see Ref.~\citep{Aliakbar})
\begin{eqnarray}\label{fit}
   u_{\max}^{(N)}\approx \frac{N^2}{2} \quad &&{\rm for \,\, even }\,\, N, \nonumber \\
   u_{\max}^{(N)}\approx \frac{N^2+1}{2} \quad &&{\rm for \,\, odd }\,\, N.
\end{eqnarray}
It has also been shown that to achieve this energy density all $N$ kinks and antikinks must collide at the same point, which is possible only if they behave as mutually attractive solitons, and thus, they should be arranged in such a way that each kink (antikink) has nearest neighbors of the opposite topological charge.

Substituting Eq. (\ref{Kink}) into Eq. (\ref{Energy}) one finds the total energy of the kink
\begin{equation}\label{KinkEnergy}
U_k=\frac{4\delta}{3}\,.
\end{equation}

To carry out numerical simulation, the discrete form of Eq.~(\ref{SG}) is considered as follows
\begin{eqnarray}\label{SG discrete}
&&\frac{d^2\phi_n}{dt^2} - \frac{1}{h^2}(\phi_{n-1} -2\phi_{n}+\phi_{n+1}) \nonumber \\
&&+\frac{1}{12h^2}(\phi_{n-2}-4\phi_{n-1} +6\phi_{n}-4\phi_{n+1}+\phi_{n+2})\nonumber \\
&&-2\phi_{n}(1-\phi_{n}^{2}) = 0,
\end{eqnarray}
in which $h$ is the lattice spacing, $n=0,\pm1,\pm2,...$, and $\phi_n(t)=\phi(nh,t)$. In order to decrease the effect of discreteness, the term $\phi_{xx}$ in Eq.~(\ref{SG discrete}) is discretized with the accuracy $O(h^4)$ \cite{BraunKivshar,DanialPRD}. The equations of motion in the form of Eq.~(\ref{SG discrete}) were integrated with respect to the time using an explicit scheme with the time step $\tau$ and the accuracy of $O(\tau^4)$. The time and the spatial steps are $h=0.05$ and $\tau=0.005$, respectively.

\begin{figure}\center
\includegraphics[width=7.5cm]{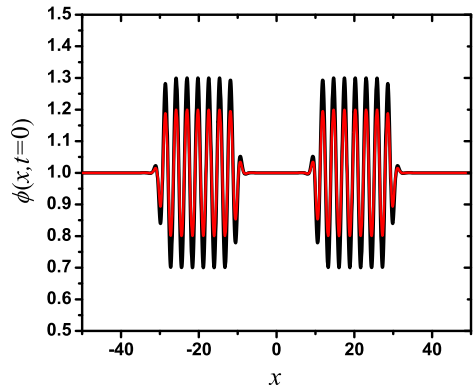}
\caption{Two examples of the incoming wave trains of Eq.~(\ref{WaveTrain}) at $t=0$ with parameters $a_1=10$, $a_2=30$, $k=2.25$ for $A=0.15$ (black line) and $A=0.1$ (red line).}\label{fig1}
\end{figure}

The initial conditions used in our simulations represent two widely separated identical wave trains propagating from both sides on the trivial background towards a collision point \cite{RomanczukiewiczPRL}:
\begin{eqnarray}\label{WaveTrain}
   \phi(x,t)=1+A[F(x+vt)\sin(\omega t+kx) +F(vt-x)\sin(\omega t-kx)],
\end{eqnarray}
where $k$ is the wave number of the incoming wave, $\omega=\sqrt{k^{2}+4}$ is the frequency and $v=k/\omega$ is the wave train velocity. We consider the envelope of the train as $F(x)=\tanh(x-a_1)-\tanh(x-a_2)$. The parameters $a_1$ and $a_2$ define the length of the train and the initial separation between the trains. We used the values $a_1=10$ and $a_2=30$. The amplitude $A$ and the wave number $k$ are the impact parameters, which can be changed freely. In Fig.~\ref{fig1} we show the wave trains of Eq.~(\ref{WaveTrain}) at $t=0$ with the initial parameters $k=2.25$, $A=0.15$ (black line) and $A=0.1$ (red line).

\begin{figure}\center
\includegraphics[width=8.5cm,height=7.0cm]{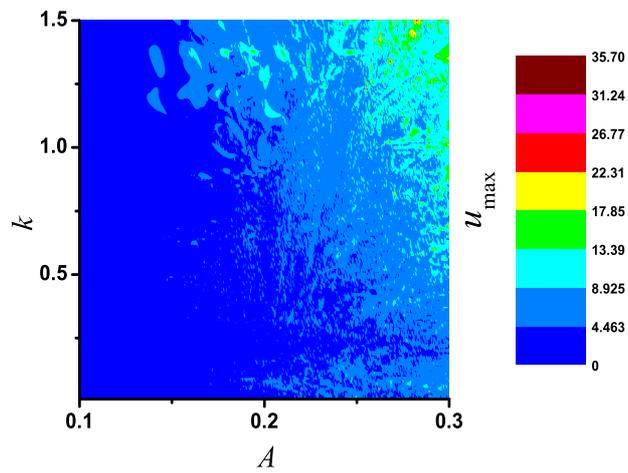}
\caption{Contour plot for the maximal energy density, $u_{\max}$, observed during collision of the two wave trains on the $(A,k)$ plane.}\label{fig2}
\end{figure}

\section {Numerical results} \label{Sec:III}

\begin{figure}\center
\includegraphics[width=8.5cm,height=7.0cm]{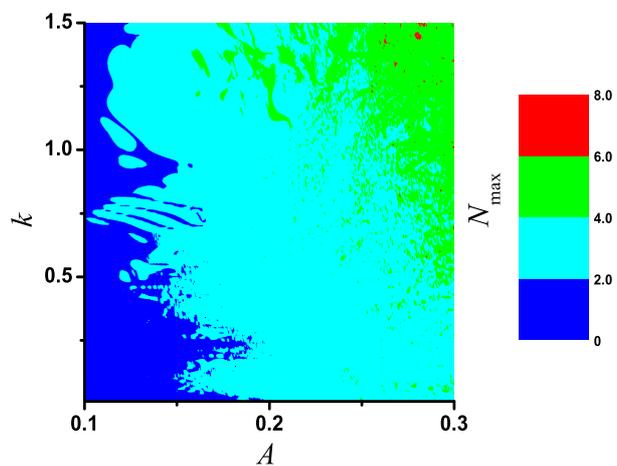}
\caption{Contour plot for the expected maximal number of kinks produced during the collision of the two wave trains with parameters $A$ and $k$.} \label{fig3}
\end{figure}
\begin{figure}\center
\includegraphics[width=8.5cm]{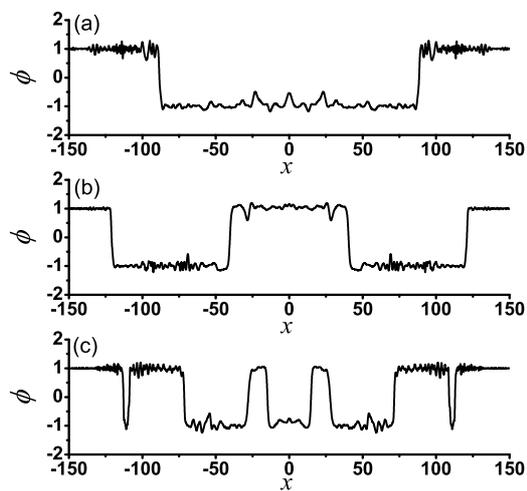}
\caption{Three snapshots of the wave train collisions with parameters (a) $k=0.58$, $A=0.285$, (b) $k=0.06$, $A=0.288$, and (c) $k=1.49$, $A=0.299$. The three graphs are plotted at $t=150$. } \label{fig4}
\end{figure}

Maximal energy density, $u_{\max}$, produced by the colliding wave trains depends on their initial parameters. The contour plot in Fig.~\ref{fig2} shows the maximal energy density observed during collision of two incoming wave trains with $a_1=10$, $a_2=30$ on the $(A,k)$ plane. It can be seen that $u_{\max}$ increases in a fractal manner with increasing $A$ and $k$.

For a certain range of impact parameters $A$ and $k$ creation of kink-antikink pairs is observed. We note that due to the conservation of topological charge only even number of kinks can be created. Creation of the kink-antikink pairs requires sufficiently large energy density. Let us use the first line of Eq.~(\ref{fit}) to predict the maximal number of kinks, $N_{\max}$, that can be formed in wave train collisions. The prediction is based on the assumption that if the maximal energy density in a wave train collision is sufficiently large to create as maximum $N$ kinks, then $N_{\max}=N$. It should be pointed out that a part of energy of the incoming wave trains is inevitably lost to radiation so that to create $N$ kinks the maximal energy density should be somewhat higher than that predicted by Eq.~(\ref{fit}). Using the results presented in Fig.~\ref{fig2} and the first line of Eq.~(\ref{fit}) we plot Fig.~\ref{fig3}, which gives the predicted maximal number of kinks that can be formed in the collision of wave trains with the parameters $A$ and $k$. Note that for $A<0.1$ and $k<1.5$ maximal energy density is small, $u_{\max}<2$, and it is insufficient to create even one kink-antikink pair.

In our numerical simulations we have found that in most cases the number of kinks created in wave train collisions can be predicted from the consideration of the maximal energy density observed in the collisions. A few examples are shown in Fig.~\ref{fig4}, where we plot the field profile for $t=150$ and for the wave trains parameters (a) $k=0.58$, $A=0.285$ (b) $k=0.06$, $A=0.288$ and (c) $k=1.49$, $A=0.299$. In (a) $u_{\max}=30.2$ so that no more than 6 kinks can be formed and in fact two kinks have emerged. In (b) $u_{\max}=15.4$ that allowes no more than four kinks to appear and this number of kinks is actually formed. In (c) $u_{\max}=30.3$ meaning that no more than 6 kinks should appear and in line with this expectation one can see 6 kinks together with two low-frequency breathers. The two breathers appear as the most remote from the origin kink-antikink pairs but in fact, they do not have enough energy to split. We have checked many more points in the $(A,k)$ parameter plane and found no violations in the prediction of the maximal number of kink-antikink pairs formed in the wave train collisions.

\section {Conclusions} \label{Sec:IV}

The simulations conducted in \cite{RomanczukiewiczPRL} have been revisited with the aim to demonstrate that the maximal number of kink-antikink pairs created in wave train collisions can be predicted with a reasonable accuracy by calculating the maximal energy density observed in the wave train collisions and comparing it with the maximal energy density observed in multi-kink collisions in \cite{Aliakbar}. 

In many cases the number of kink-antikink pairs created in wave train collisions was smaller than the predicted maximal possible number. As an example, refer to Fig.~\ref{fig4}(a) where maximal possible number of kinks is six but only two were created. We believe that this is because there is one more important parameter defining the initial conditions, which is the initial phase of the incoming wave trains. Only mirror-symmetric colliding wave trains were considered, which severely restricts dynamics of the system.

\section*{Acknowledgments}

S.V.D acknowledges financial support provided by the Russian Science Foundation, grant 16-12-10175.



\bibliographystyle{elsarticle-num}

\bibliography{sample}

\end{document}